\documentclass[aps,prd,twocolumn,groupedaddress,showpacs]{revtex4}

\usepackage{graphicx}

\begin{document}


\title{A Simultaneous Solution to the $^6$Li and $^7$Li Big Bang Nucleosynthesis Problems from a Long-Lived Negatively-Charged Leptonic Particle}


\author{Motohiko Kusakabe$^{1,2}$\footnote{Research Fellow of the Japan
Society for the Promotion of Science}}
\email[]{kusakabe@th.nao.ac.jp}

\author{Toshitaka Kajino$^{1,2,3}$, Richard N. Boyd$^{1,4}$, Takashi
Yoshida$^{1}$ and Grant J. Mathews$^{5}$}

\affiliation{
$^1$Division of Theoretical Astronomy, National Astronomical Observatory
of Japan, Mitaka, Tokyo 181-8588, Japan \\
$^2$Department of Astronomy, Graduate School of Science, University of
Tokyo,  Hongo, Bunkyo-ku, Tokyo 113-0033, Japan \\
$^3$Department of Astronomical Science, The Graduate University for
Advanced Studies, Mitaka, Tokyo 181-8588, Japan \\
$^4$Lawrence Livermore National Laboratory, University of California,
Livermore, CA 94550, USA \\
$^5$Department of Physics and Center for Astrophysics, University of
Notre Dame, Notre Dame, IN 46556, USA }


\date{\today}

\begin{abstract}

The $^6$Li abundance observed in metal poor halo stars exhibits a plateau
 similar to that for $^7$Li suggesting a primordial origin.  However, the
 observed abundance of $^6$Li is a factor of $10^3$ larger and that of
 $^7$Li is a factor of 3 lower than the abundances predicted in the standard
 big bang when the baryon-to-photon ratio is fixed by WMAP.  Here we
 show that both of these abundance anomalies can be explained by the
 existence of a long-lived massive, negatively-charged leptonic particle
 during nucleosynthesis.  Such particles would capture onto the synthesized nuclei thereby reducing the reaction Coulomb barriers and
 opening new transfer reaction possibilities, and catalyzing a
 second round of big bang nucleosynthesis. This novel solution to both
 of the Li problems can be achieved with or without the additional
 effects of stellar destruction.

\end{abstract}



\maketitle

\section{Introduction}

It has recently been pointed out (e.g.,~\cite{Asplund:2005yt})  that the
abundances of both $^6$Li and $^7$Li observed in metal poor halo stars
(MPHSs) are not in agreement with those predicted from standard big-bang
nucleosynthesis (BBN). Specifically, the $^6$Li abundance as a function
of metallicity exhibits a plateau similar to that for $^7$Li in very
metal-poor stars, suggesting a primordial origin for both isotopes.
This $^6$Li abundance plateau, however, is a factor of $\sim 10^3$
larger than that predicted by BBN.  A less
severe problem exists for $^7$Li; the BBN value based upon the
baryon-to-photon ratio fixed by the WMAP analysis~\cite{Spergel:2006hy} of the cosmic microwave background is roughly a factor
of three higher than is observed. 
 
Moreover, a longstanding effort in cosmology has been the search for
     evidence of unstable particles that might have existed in the early
     universe.  It is thus natural to ask whether the two lithium
     abundance anomalies might be a manifestation of the existence of
     such a particle.  In this context a number of possible solutions to
     the $^6$Li problem have been proposed which relate the Li anomalies
     to the possible existence of unstable particles in the early
     universe~\cite{decaying}.  This has been extended in several recent
     studies~\cite{Pospelov:2006sc,Cyburt:2006uv,Kaplinghat:2006qr,Kohri:2006cn,Bird:2007ge,Hamaguchi:2007mp}
     to consider heavy negatively charged decaying particles that modify BBN, but in
     rather different ways.  In these latter studies, the heavy particles,
     here denoted as $X^-$, bind to the nuclei produced in BBN to form
     $X$-nuclei.  The massive $X^-$ particles would be bound in orbits
     with radii comparable to those of normal nuclei.  Hence, they would
     reduce the reaction Coulomb barriers thereby enhancing the
     thermonuclear reaction rates and extending the duration of BBN to
     lower temperatures.

     Pospelov~\cite{Pospelov:2006sc} suggested that a large enhancement of the $^6$Li abundance could result
     from a transfer reaction involving an $X^-$ bound to $^4$He (denoted
     $^4$He$_X$), i.e. $^4$He$_X$($d$,$X^-$)$^6$Li.  Although this was
     an intriguing idea, Hamaguchi et al.~\cite{Hamaguchi:2007mp} have
     recently pointed out via a more complete quantum mechanical calculation that Pospelov's estimate for the $^4$He$_X$($d$,$X^-$)$^6$Li cross section was too large; this leads to too high a $^6$Li abundance.  Cyburt et al.~\cite{Cyburt:2006uv} further motivated this
     hypothesis by identifying the $X^-$ as the supersymmetric partner
     of the tau lepton, i.e., a stau, and considered $X^-$ transfer
     reactions for $^7$Li and $^7$Be production, too.  Although their
     calculation is based on a fully dynamical treatment of the
     recombination of nuclei with $X^-$ particles and also of BBN, they
     used cross sections for all $X^-$ transfer reactions involving $^6$Li, $^7$Li, and $^7$Be that were too large, as we discuss below.  Therefore, the
     calculated abundances are to be viewed with caution.  Kaplinghat and
     Rajaraman~\cite{Kaplinghat:2006qr} observed that the decay of an
     $X^-$ when bound to $^4$He would occasionally knock out a proton or
     neutron to produce $^3$He or $^3$H, thereby enhancing their
     abundances and the abundance of $^6$Li through reactions with other
     primordial $^4$He nuclei at higher energies.  Kohri and
     Takayama~\cite{Kohri:2006cn} studied the recombination of nuclei
     with $X^-$ particles, and suggested the possibility of solving the
     $^7$Li problem.  However, they did not carry out dynamical
     calculations involving recombination processes and BBN
     simultaneously.  This forced them to introduce artificial
     parameters for the fractions of the captured nuclei, which turn out
     to be different from the fractions obtained by solving the
     recombination plus BBN fully dynamically.  A new resonant reaction
     $^7$Be$_X$($p$,$\gamma$)$^8$B$_X$ has recently been proposed by
     Bird et al.~\cite{Bird:2007ge} that destroys
     $^7$Be$_X$ through an atomic excited state of $^8$B$_X$, and the
     present study identifies another effect in this reaction that might
     also destroy $^7$Be. Thus there has been a great deal of recent
     progress in $X^-$ catalyzed BBN in three important aspects: the
     simultaneous description of recombination and ionization processes
     of $X^-$ particles with nuclei in the description of BBN, use of updated reaction rates involving the $X$- nuclei, and inclusion of new resonant processes by which $^7$Be is destroyed. No previous calculation has involved all of these effects in a single dynamical calculation of BBN in order to study their effects on the $^6$Li and $^7$Li problems.

     In this Article, we present the results of a thorough dynamic analysis
     of the effects of $X^-$ particles on BBN.  The important difference
     from previous works is, firstly, that we carried out a fully dynamical BBN
     calculation by taking account of the recombination and ionization
     processes of $X^-$ particles by normal and $X$-nuclei as well as the
     many possible nuclear reactions among them.  Secondly, the reaction rates on normal and $X$-nuclei used in the present study are based on quantum mechanical
     calculations of the cross section like those of Hamaguchi et
     al.~\cite{Hamaguchi:2007mp}, which we believe to be correct.  Thirdly, we have not
 only included the important $^7$Be destruction mechanism identified by Bird et
 al.~\cite{Bird:2007ge}, but have identified another potentially important destruction mechanism involving the reaction channel $^7$Be$_X$+$p$ $\rightarrow ^8$B$^*$($1^+$, 0.770 MeV)$_X$
     $\rightarrow ^8$B$_X$+$\gamma$ which has the potential to destroy $^7$Be$_X$ via capture through
     the $1^+$ nuclear excited state of $^8$B.  We show that when all
     these effects are included, the single hypothesis of the existence
     of the $X^-$ particle provides a remarkable solution to both the $^6$Li and $^7$Li abundance anomalies.  We can then use this constraint to place interesting limits on the $X^-$ relic abundance
     and its decay lifetime and mass.

\section{Model}\label{sec2}
We assume that the $X^-$ particle is
leptonic and of spin 0, consistent with its identification as
the supersymmetric partner of a lepton. The $X^-$ would be thermally
produced at an earlier epoch together with $X^+$.  Their small
annihilation cross section allows a significant abundance to
survive to the BBN epoch. The mass and decay lifetime of the $X^-$ is
ultimately constrained by WMAP and the present BBN study.  Only the
$X^-$ can bind to nuclei and the $X^+$ remains inert during BBN. The
binding energies and the eigenstate wave functions of the $X$-nuclei were
calculated by assuming uniform finite-size charge
distributions of radii $r_0=1.2 A^{1/3}$~fm for nuclear mass number $A$~\cite{cahn:1981}.  When the $X^-$ abundance is very high,
some nuclei can bind two $X^-$ particles, such as $^3$He$_{XX}$ and
$^4$He$_{XX}$. In that case their binding energies were calculated using
a variational calculation with a trial wave function for $X$-nuclides
bound to one $X^-$ particle, analogous to the case of the H$_2^+$ ion.

Thermonuclear reaction rates (TRRs) for all reactions that might take place in $X^-$ catalyzed BBN,
     including the $X^-$ transfer reaction suggested in~\cite{Pospelov:2006sc}
     and $X^-$ decay, were added to the BBN network code.  See Kusakabe
     et al.~\cite{kusakabe} for details on the calculations.  These were
     corrected for the modified nuclear charges and the effective mass
     resulting from the binding of one or two $X^-$ particle(s).  If the
     $X^-$ decayed at some later stage, they would be expected to
     destroy some fraction of the nuclei to which they had become bound
     during BBN.  However, that fraction would be small~\cite{Kaplinghat:2006qr,rosen:1975}.  We found that the
     inclusion of the $X$-nuclei $^8$Be$_X$ and $^8$Be$_{XX}$ (both are bound) results in a leakage of the nuclear reaction flow
     out of the light nuclei or $X$-nuclei to produce slightly heavier
     $A \geq$~9 nuclei. This might be an additional BBN signature
     resulting from binding the $X^-$ particles.  We determined most
     thermonuclear reaction rates involving  the $X$-nuclei by taking account of
     the lowered Coulomb barriers and modified reduced masses.  However,
     as discussed below, there are a number of reactions that require
     careful additional considerations.
     
     As noted by Pospelov~\cite{Pospelov:2006sc}, reactions in which an $X^-$ particle is
     transferred can be very important in circumventing some normally
     inhibited reactions, especially the
     $^4$He$_X$($d$,$X^-$)$^6$Li  reaction. Its rate could be orders of
     magnitude larger than that of the $^4$He($d$,$\gamma$)$^6$Li
     reaction, which is suppressed due to its occurrence through an
     electric quadrupole transition.  Hamaguchi et
     al.~\cite{Hamaguchi:2007mp} have recently carried out a theoretical
     calculation of the cross section for $^4$He$_X$($d$,$X^-$)$^6$Li in a
     quantum three-body model.  Their value was about an order of
     magnitude smaller than that of~\cite{Pospelov:2006sc}.  This
     difference can be attributed to the use of an exact treatment of
     quantum tunneling and a better nuclear potential.  We, therefore,
     adopt the result of~\cite{Hamaguchi:2007mp} in the present study.

     Cyburt et al.~\cite{Cyburt:2006uv} estimated astrophysical S-factors for the
     $^4$He$_X$($t$,$X^-$)$^7$Li, $^4$He$_X$($^3$He,$X^-$)$^7$Be,
     $^6$Li$_X$($p$,$X^-$)$^7$Be and other reactions by applying a scaling
     relation~\cite{Pospelov:2006sc}, $S_X$/$S_\gamma \propto p_{\rm
     f}a_0/(\omega_\gamma a_0)^{2\lambda+1}$.  Here, $S_X$ and
     $S_\gamma$ are the S-factors for the $X^-$ transfer and radiative
     processes, respectively, $a_0$ is the $X^-$ Bohr radius of
     $^4$He$_X$ or $^6$Li$_X$, $p_{\rm f}$ is the linear momentum of the
     outgoing $^7$Li or $^7$Be in the $X^-$ transfer reactions, and
     $\omega_\gamma$ is the energy of  the emitted $\lambda=1$ (electric
     dipole) photon in the radiative capture. However, the reaction
     dynamics are important to these results.
     
     $^4$He, $^{6,7}$Li, and $^7$Be occupy an s-wave orbit around the
     $X^-$ particle (assuming the $X^-$ particle to be much heavier than
     these nuclei).  The $^6$Li nucleus is an $\alpha$+$d$ cluster system
     in a relative s-wave orbit, while the $A=7$ nuclei are $\alpha$+$t$ and
     $\alpha$+$^3$He cluster systems in relative p-wave orbits. This
     difference in the orbital angular momentum will produce a critical
     difference in the reaction dynamics between the
     $^4$He$_X$($d$,$X^-$)$^6$Li and the $^4$He$_X$($t$,$X^-$)$^7$Li,
     $^4$He$_X$($^3$He,$X^-$)$^7$Be, and $^6$Li$_X$($p$,$X^-$)$^7$Be
     reactions.  Specifically, the latter three reactions must involve $\Delta l$=1
     angular momentum transfer. In order to conserve total angular
     momentum the outgoing $^7$Li and $^7$Be in the final state must
     therefore occupy a scattering p-wave orbit from the $X^-$
     particle, leading to a large hindrance of the overlap matrix for
     the $X^-$ transfer processes. Thus, a realistic quantum mechanical
     calculation results in much smaller $S_X$-factors than those
     estimated in~\cite{Cyburt:2006uv}.  Therefore, in
     the present study, the above three reaction processes were found to be negligible and were therefore omitted. 

     Bird et al.~\cite{Bird:2007ge} suggested that the
     $^7$Be$_X$($p$,$\gamma$)$^8$B$_X$ resonant reaction could destroy
     $^7$Be$_X$ through an atomic excited state of $^8$B$_X$.  They also
     proposed that a charged weak-boson exchange reaction
     $^7$Be$_X \rightarrow ^7$Li+$X^0$ followed by $^7$Li($p$,$\alpha$)$^4$He could destroy $A=7$ nuclides.  We included only the former resonant
     reaction in the present study, although we confirmed their
     assertion on the weak process as will be discussed in a separate
     paper.

     In our exhaustive study of additional processes related to
     $^6$Li, $^7$Li, and $^7$Be destruction, we found that the reaction channel which proceeds 
     through the $1^+$, $E^*=0.770\pm 0.010$ MeV nuclear excited state
     of $^8$B via $^7$Be$_X$+$p$ $\rightarrow ^8$B$^*$($1^+$, 0.770
     MeV)$_X$ $\rightarrow ^8$B$_X$+$\gamma$ could also destroy some
     $^7$Be$_X$, and that the destruction processes
     $^6$Li$_X$($p$,$^3$He)$^4$He$_X$, $^7$Li$_X$($p$,$\alpha$)$^4$He$_X$
     might also be significant.  

Our calculated binding energies of the $X^-$ particle in
$^7$Be$_X$ and $^8$B$_X$ are respectively 1.488 MeV and 2.121 MeV.
Adopting these values without any correction to the energy levels of the
nuclear excited states of $^8$B$_X$, this $1^+$ state of $^8$B$_X$
is located near the particle threshold for the $^7$Be$_X$+$p$ separation
channel.  Thus, the $^7$Be$_X$($p$,$\gamma$)$^8$B$_X$ reaction can
proceed through a zero-energy resonance of $^8$B$^\ast_X$.  However, the
measured energy uncertainty of the $1^+$ state of $^8$B is $\pm 10$~keV, and moreover, the excitation energy of this level is very sensitive to the
model parameters used to calculate the binding energies of the $X$-nuclei.  Even
such a small uncertainty of the resonance energy as 10-100~keV
would dramatically change the TRR because the BBN catalyzed by the
$X$-nuclei proceeds at effective temperatures as low as $T_9\sim 0.1$.
Taking account of the uncertainties associated with the $1^+$ resonance energy, $E$, from the $^7$Be$_X$+$p$ separation
threshold, we found that $E \approx 30$~keV maximizes the TRR.  This
threshold energy would be achieved when, for example, the uniform charge radii
are 2.2955~fm for $^7$Be$_X$ and 2.4564~fm for $^8$B$_X$, respectively.
This resonant reaction is potentially as effective as
 $^7$Be$_X$+$p$ $\rightarrow$ $^8$B$_X^{*a} \rightarrow
^8$B$_X+\gamma$ in destroying $^7$Be.  However, the
charge radii we have adopted tend to be smaller than the measured charge radii, and this 
might overestimate the binding energies of $X^-$.  If a more realistic
calculation were performed, the resulting binding energies might shift
this $1^+$ excited state upward, which would diminish the effect of this destruction process.  In addition, the transition through this state would be E2 or M1, which might also weaken its effect. Even in this case, though, the atomic resonance
$^8$B$_X^{*a}$~\cite{Bird:2007ge} plays the important role in destroying
$^7$Be$_X$.

  Since
     it is important to know precisely when during BBN the $X^-$ particles become bound
     to nuclei, and what their distribution over the BBN nuclei would
     be~\cite{Kohri:2006cn} at any time it is necessary to consider the
     thermodynamics associated with binding the $X^-$ particles.  
     We thus included both recombination and
     ionization processes for $X^-$ particles in our BBN network code
     and dynamically solved the set of rate equations to find when the $X$-nuclei decoupled
     from the cosmic expansion.

     Regarding the thermonuclear reaction rates we note that since the
     mass of $X^-$ particle $m_X$ is assumed to be $\gtrsim$ 50 GeV, the
     reduced mass for the $X^-$+$A(N,Z)$ system can be approximated as
     $\mu_X \equiv m_A m_X/(m_A+m_X) \approx m_A$, rendering the
     thermonuclear reaction rate for the first recombination process
     $A(X^-,\gamma)A_X$~\cite{Kohri:2006cn} to be $\langle \sigma_r
     v\rangle_X \approx 2^9 \pi \alpha Z^2 (2\pi)^{1/2}/(3\exp(4.0)) E_{\rm
     bind}/(\mu_X^2 (\mu_X T)^{1/2}) \propto m_A^{-2.5}$, where $\alpha$
     is the fine structure constant.  This rate is almost independent of
     $m_X$.  However, the rate for the second recombination process
     $A_X(X^-,\gamma)A_{XX}$ is dependent upon $m_X$, i.e., $\langle \sigma_r
     v\rangle_{XX} \approx 2^9 \pi \alpha (Z-1)^2 (2\pi)^{1/2}/(3\exp(4.0))
     E_{\rm bind}/(\mu_{XX}^2 (\mu_{XX}T)^{1/2}) \propto m_X^{-2.5}$.
     This arises because $\mu_{XX} \equiv m_{AX} m_X/(m_{AX}+m_X)
     \approx m_X/2$.  Since $m_X$ is assumed to be much larger than the
     mass of the light nuclei $m_X \gg m_A$, the rate for the second or
     higher-order recombination process is hindered.
		
\section{Results}\label{sec3}
The evolution of the BBN abundances when $X^-$
particles are included exhibits some particularly notable
features. During the nucleosynthesis epoch, the abundances for $^6$Li,
$^7$Li and $^7$Be assume their normal BBN values until the temperature
reaches $T_9 \sim 0.5-0.2$.  Below that temperature the $X^-$ particles
bind to the heaviest nuclides, $^7$Li and $^7$Be. When the abundance
ratio, $Y_X$,  of $X^-$ particles to baryons is larger than 0.1   these
nuclides are then partially destroyed by reactions that would have
previously been inhibited by the Coulomb barrier.  At around $T_9 =
0.1$, the $X^-$ particles are captured onto $^4$He.  Then a new round of
$X$-nuclei nucleosynthesis occurs. In particular, the reaction
$^4$He$_X$($d$,$X^-$)$^6$Li produces normal $^6$Li nuclei with an abundance
which is orders of magnitude above that from standard BBN.  An
interesting feature is that the $^6$Li formed in this way is not easily destroyed by the
$^6$Li($p$,$\alpha$)$^3$He reaction, the dominant $^6$Li destruction reaction in BBN, because the $X^-$ transfer reaction
restores the charge to $^6$Li.  Hence, the Coulomb barrier is too high
at this temperature for its destruction resulting in a large
$^6$Li/$^7$Li abundance ratio.

The final calculated abundances of the mass 6 and 7 nuclides, however,
depend strongly on the assumed $X^-$ abundance.  At high $X^-$ abundance
levels, more than one $X^-$ capture can occur.  Although the abundance
of these multiple $X^-$ bound particles is too small to significantly
contribute to BBN, they nevertheless interact readily since their
Coulomb barriers are greatly reduced. This is especially true of
charge-neutral $^4$He$_{XX}$.  To clarify the nucleosynthesis yields, we
have thus made a study in which the $X^-$ abundance $Y_X$ was varied
over a wide range.

In Fig. 1 we show contours of an interesting region in  the decay lifetime $\tau_X$ vs. $Y_X$
plane.  Curves are drawn for constant lithium abundance relative to the
observed value in MPHSs, i.e., d($^6$Li) = $^6$Li$^{\rm
Calc}$/$^6$Li$^{\rm Obs}$ (solid curves)  and d($^7$Li) = $^7$Li$^{\rm
Calc}$/$^7$Li$^{\rm Obs}$  (dashed curves) for several values of the
stellar depletion factor ^^ ^^ d''.  The adopted abundances are
$^7$Li/H$=(1.23^{+0.68}_{-0.32})\times 10^{-10}$~\cite{Ryan:1999vr} and
$^6$Li/H$=(7.1\pm 0.7)\times 10^{-12}$~\cite{Asplund:2005yt}.  Shaded
regions for the d($^6$Li) = 1  and d($^7$Li) = 1 curves illustrate the
1~$\sigma$ uncertainties in the adopted observational constraints based
upon the dispersion of the observed plateaus. We also show curves for
stellar depletion factors  of d($^7$Li) = 2, 3 and d($^6$Li) = 4, 25.  
Since $^6$Li is more fragile to stellar processing than $^7$Li
~\cite{Richard:2004pj}, its possible depletion factors could be
 larger than those for $^7$Li.  
 
The main point of this figure is
that, independent of stellar destruction, it is possible to find a
simultaneous solution to both the $^7$Li overproduction problem and the
$^6$Li underproduction problem. This occurs in the parameter region $Y_X
\approx 0.09-0.6$, $\tau_X \approx (1.6-2.8)\times 10^3$~s consistent
with the suggestion of~\cite{Bird:2007ge}.  Assuming that the products
of the decaying $X^-$ particles are progenitors of the CDM particles,
the WMAP-CMB observational constraint on $\Omega_{\rm CDM}$ = 0.2 limits the
mass of the $X^-$, i.e., $Y_X m_X \lesssim$ 4.5 GeV and $m_X \lesssim$ 50 GeV
when we include the destruction reaction processes of $A=7$ nuclide
$^7$Be$_X$+$p$ $\rightarrow$ $^8$B$_X^{*a} \rightarrow
^8$B$_X+\gamma$~\cite{Bird:2007ge} and (assumed maximal value of the) $^7$Be$_X+p\rightarrow
^8$B$^*$($1^+$, 0.770 MeV)$_X \rightarrow ^8$B$_X+\gamma$.  When we
include the destruction process $^7$Be$_X \rightarrow
^7$Li+$X^0$~\cite{Bird:2007ge}, these parameter ranges slightly change
to $Y_X \approx 0.04-0.1$, $\tau_X \approx (1.8-3.2)\times 10^3$~s, and
$m_X \lesssim 100$~GeV.  Figure 2 illustrates the final calculated BBN
yields as a function of baryon to photon ratio $\eta$ for the case of
($Y_X$, $\tau_X$)=(0.6, $1.6\times 10^3$~s).  This choice leads to
$^6$Li and $^7$Li abundances consistent with the observed values without stellar depletion.  Note, though, that the same conclusion is reached if the destruction reaction through the $^8$B($1^+$) state is not included, so the general conclusion is robust.

\begin{figure}[tbp]
\includegraphics[width=8.0cm,clip]{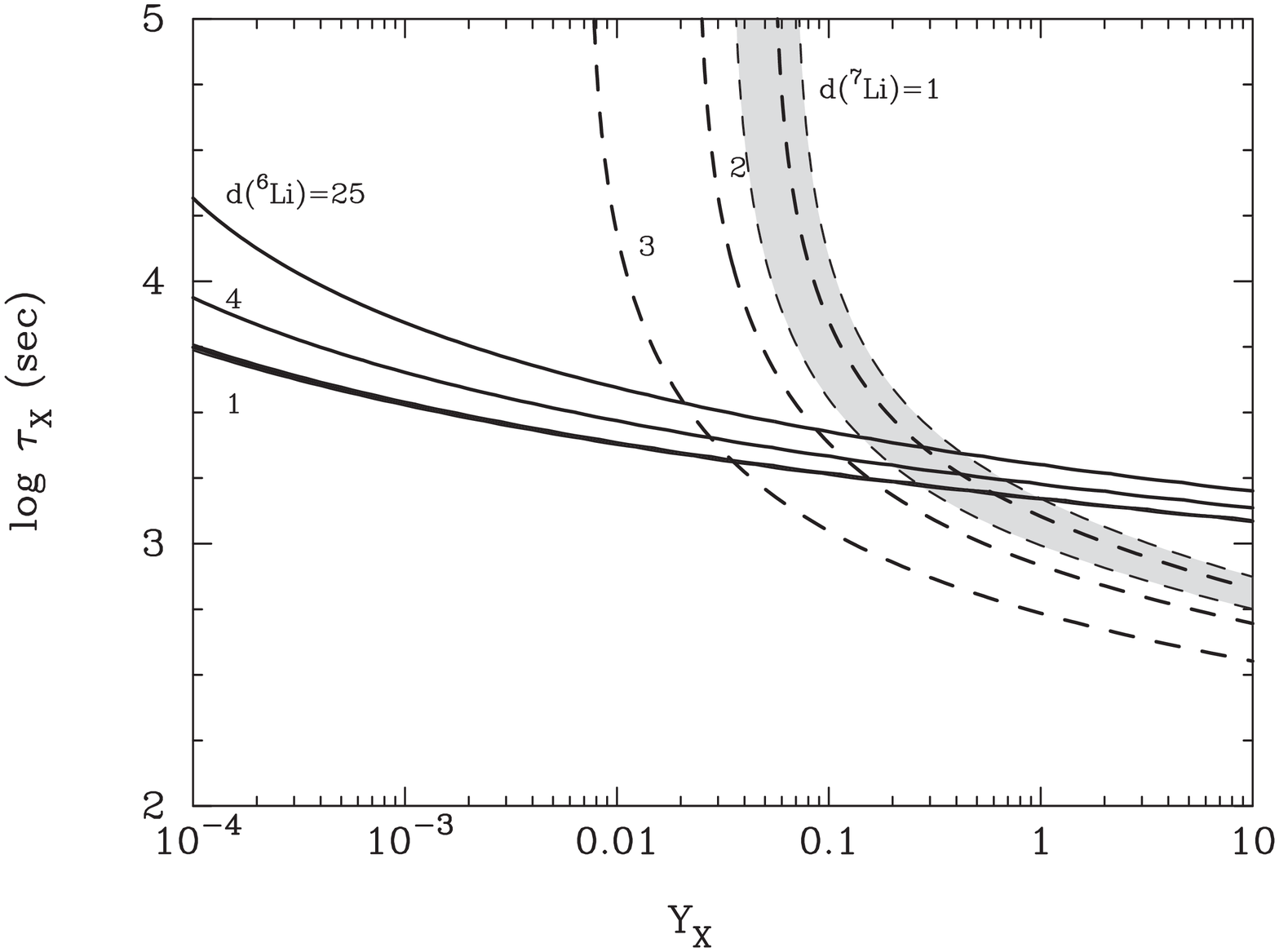}
\caption{\label{contour} Contours of constant lithium abundance relative
 to the observed value in MPHSs, i.e., d($^6$Li) = $^6$Li$^{\rm Calc}$/$^6$Li$^{\rm Obs}$ (solid curves) and d($^7$Li) = $^7$Li$^{\rm Calc}$/$^7$Li$^{\rm Obs}$ (dashed curves).  The adopted abundances are $^7$Li/H$= (1.23^{+0.68}_{-0.32})\times 10^{-10}$~\cite{Ryan:1999vr} and $^6$Li/H$=(7.1\pm 0.7)\times 10^{-12}$~\cite{Asplund:2005yt}.  Shaded regions for the d($^6$Li) = 1  and d($^7$Li) = 1 curves illustrate the 1~$\sigma$ uncertainties in the adopted observational constraints based upon the dispersion of the observed plateaus.}
\end{figure}



\begin{figure}[tbp]
\includegraphics[width=8.0cm,clip]{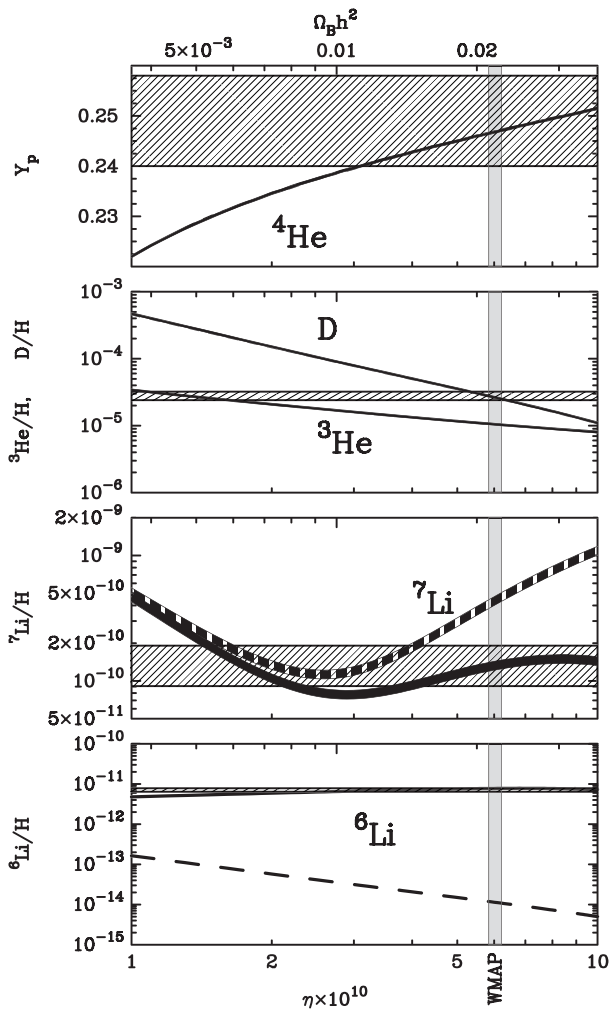}
\caption{\label{bbn} Abundances of $^4$He (mass fraction), D, $^3$He,
 $^7$Li and $^6$Li (by number relative to H) as a function of the
 baryon-to-photon ratio $\eta$ or $\Omega_B h^2$.  The dashed and solid
 curves are respectively the calculated results in the standard BBN and
 the $X^-$ catalyzed BBN for the case of ($Y_X$, $\tau_X$)=(0.6,
 $1.6\times 10^3$s).  There is virtually no difference between the
 dashed and solid curves for $^4$He, D, and $^3$He.  The band of
 theoretical curve for each nucleus displays 1~$\sigma$ limits taken
 from~\cite{Coc:2003ce}.  The hatched regions represent the adopted
 abundance constraints from~\cite{oli04} for
 $^4$He,~\cite{Kirkman:2003uv} for D,~\cite{Ryan:1999vr} for $^7$Li,
 and~\cite{Asplund:2005yt} for $^6$Li, respectively.  The vertical
 stripe represents the 1~$\sigma$~$\Omega_B h^2$ limits provided by
 WMAP~\cite{Spergel:2006hy}.}
\end{figure}


\section{Summary}\label{sec4}
In summary, we have investigated light-element nucleosynthesis during
BBN taking into account the possibility of massive, negatively-charged
$X^-$ particles which would bind to the light-nuclei.  When the chemical
and kinetic processes associated with such particles are included in a BBN code in a fully
dynamical manner, along with the reactions enabled by the $X^-$
particles, the $X^-$ particles are found to enhance the reaction rates
in BBN, both by reducing the charge of the resulting $X$-nuclei, and by
enabling transfer reactions of the $X^-$ particles.  $X^-$ particles
greatly enhance the production of $^6$Li, primarily from the $X^-$
transfer reaction $^4$He$_X$($d$,$X^-$)$^6$Li.  The $^7$Li abundance,
however, decreases when the $X^-$ particle abundance is larger than~0.1
times the total baryon abundance.  In this case, the $^7$Li abundance
decreases with the $X^-$ particle abundance due to the inclusion of two resonance
channels for  $^7$Be$_X$($p$,$\gamma$)$^8$B$_X$ through the nuclear and
atomic excited states of $^8$B$_X$.  It was found to be
important to predict precisely the binding energies and excited
states of exotic $X$-nuclei in realistic quantum mechanical
calculations.  Both abundance ratios
of $^6$Li/H
and $^7$Li/H observed in MPHSs are obtained with an appropriate choice
for the lifetime and abundance of the  $X^-$ particle.  These
observational constraints imply a lifetime and abundance roughly in the range
of $\tau_X \sim 2 \times 10^3$~s and $Y_X \sim 0.1$.
  We deduce that this $Y_X$ value requires that
$m_X \sim 50$~GeV in order to guarantee that this abundance of $X^-$ particles
survives to the epoch of nucleosynthesis.

\begin{acknowledgments}
We are very grateful to Professor Masayasu Kamimura for enlightening
 suggestions on the nuclear reaction rates for transfer and radiative
 capture reactions.  This work has been supported in part by the Mitsubishi Foundation, the
 Grant-in-Aid for Scientific Research (17540275) of the Ministry of Education,
 Science, Sports and Culture of Japan, and JSPS Core-to-Core Program of
 International Research Network for Exotic Femto Systems (EFES).  MK acknowledges the support by the Japan
 Society for the Promotion of Science.  Work at the University of Notre
 Dame was supported by the U.S. Department of Energy under Nuclear
 Theory Grant DE-FG02-95-ER40934. RNB gratefully acknowledges the
 support of the National Astronomical Observatory of Japan during his stay
 there.

\end{acknowledgments}



\end{document}